\documentclass[prl,a4paper,twocolumn]{revtex4}
\usepackage{graphicx}
\usepackage{epstopdf}
\DeclareGraphicsExtensions{.eps,.pdf}
\usepackage{float}
\usepackage[latin1]{inputenc}
\usepackage{natbib}

\begin{document}
\title{Breakdown of the Korringa Law of Nuclear Spin Relaxation in Metallic GaAs}

\author{Dominikus K\"{o}lbl}
\author{Dominik M. Zumb\"{u}hl}
\email{dominik.zumbuhl@unibas.ch} \affiliation{Department of Physics, University of Basel, Klingelbergstrasse 82,
CH-4056 Basel, Switzerland}
\author{Andreas Fuhrer}
\author{Gian Salis}
\author{Santos F. Alvarado}
\affiliation{IBM Research, Z\"{u}rich Research Laboratory, S\"{a}umerstrasse 4, 8803 R\"{u}schlikon, Switzerland}
\date{\today}

\begin{abstract}
We present nuclear spin relaxation measurements in GaAs epilayers using a new pump-probe technique in all-electrical,
lateral spin-valve devices. The measured $T_1$ times agree very well with NMR data available for $T>$\,1\,K. However,
the nuclear spin relaxation rate clearly deviates from the well-established Korringa law expected in metallic samples
and follows a sub-linear temperature dependence $T_1^{-1} \propto T^{0.6}$ for 0.1 K $\leq T \leq$ 10 K. Further, we
investigate nuclear spin inhomogeneities.
\end{abstract}
\maketitle

The coupling between the electronic and nuclear spin systems in condensed matter is of fundamental importance, leading
to many interesting effects including dynamic nuclear polarization (DNP) \cite{Lampel1969, Paget1977, Meier},
Overhauser fields \cite{Overhauser1953} as well as Knight shifts \cite{Knight1949}. The Overhauser fields can induce
electron spin decoherence but can also be exploited for coherent electron spin manipulation -- relevant in spintronics
\cite{Datta1990,Wolf2001} and quantum computation \cite{Loss1998,Hanson2007}. The nuclear spin system, on the other
hand, is likewise affected by the electrons, e.g. by the hyperfine field and nuclear-electron spin flip-flops,
contributing to nuclear spin polarization and relaxation.

In metallic systems, the small nuclear Zeeman splitting restricts the electrons participating in flip-flops to the
thermally broadened Fermi-edge, resulting in a nuclear spin relaxation (NSR) rate $T_1^{-1}$ proportional to the
electronic temperature $T$ -- the Korringa law of nuclear spin relaxation \cite{Korringa1950}. This NSR law holds for
temperatures $T$ smaller than the electronic Fermi temperature but exceeding the nuclear Zeeman splitting and further
assumes a free electron model and a dominant Fermi-contact interaction. The Korringa law has been confirmed over many
years in numerous experiments in a wide range of metals \cite{Abragam, Slichter,Anderson1959} as well as metallically
doped semiconductors \cite{Sundfors1964,Tunstall1979,Kaur2010} and is well established as the preeminent law of NSR in
metallic systems at low temperatures. As an application, the Korringa law provides the crucial link for cooling the
electronic degree of freedom in nuclear demagnetization refrigeration \cite{Lounasmaa1974,Pickett1988}. Deviations from
the Korringa law have been reported in samples at the metal-insulator transition (MIT) showing non-metallic
conductivity \cite{Paalanen1985} or in various exotic materials.

In this Letter, we report the breakdown of the Korringa law in n-doped GaAs epilayers displaying metallic conductivity.
NSR is measured with a novel pump-probe technique in lateral, all-electrical spin-valve devices
\cite{Johnson1985,Jedema2002} on GaAs \cite{Lou2007,Salis2009,Ciorga2009}, making easily accessible the low temperature
regime $T\ll 1\,\mathrm{K}$ which was not previously explored. This technique is in principle applicable to any
spin-valve device. The measured $T_1$ times agree well with NMR experiments available for high temperatures $T>
1\,\mathrm{K}$ \cite{Lu2006,Kaur2010}. The temperature dependence of the NSR rate follows a power law $T_1^{-1}\propto
T^{0.6\pm0.04}$ over two orders of magnitude in temperature $0.1\,\mathrm{K}\leq T\leq 10\,\mathrm{K}$, deviating
substantially from the Korringa law $T_1^{-1}\propto T$ for the present doping a factor of $\sim2.5$ above the GaAs MIT
well on the metallic side. The observed NSR power law $\propto T^{0.6}$ is qualitatively consistent with the combined
effects of disorder and electron-electron interactions \cite{AAreview,Miranda2005} within a hyperfine-mediated NSR
mechanism applicable here, though an appropriate theory is not currently available. At low $T$, relatively strong
coupling and correspondingly fast NSR rates are found, potentially enhancing electron cooling in nuclear refrigeration
schemes. Finally, we investigate effects of nuclear spin inhomogeneities.

\begin{figure}[tpb]
\includegraphics[width=8.5cm]{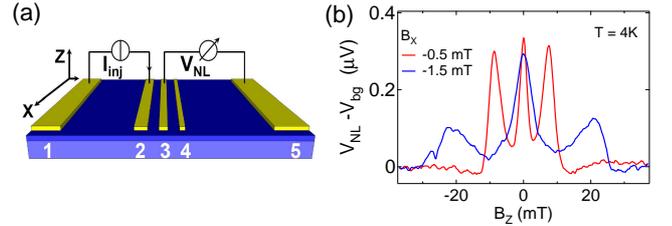}
\caption{\label{fig:1}(a) Illustration of spin-valve device and measurement setup. (b) Hanle measurements at 4\,K with
satellite peaks indicating the nuclear Overhauser field $B_N$. The non-local voltage $V_{NL}$ is shown as a function of
perpendicular field $B_Z$ (ramp rate 0.34\,mT/s) for $B_X$ as labeled. A parabolic background $V_{bg}(B_Z)$ has been
subtracted.}
\end{figure}

The spin-valves, shown in Fig.\,\ref{fig:1}(a), consist of 6\,nm thick Fe bars on a c(4x4) reconstructed surface of a
$1\,\mathrm{\mu m}$ thick GaAs epilayer with carrier density $n=5\,\times\,10^{16}\,\mathrm{cm}^{-3}$. A 15\,nm thick,
much higher doped GaAs surface layer ensures efficient spin injection. The center contacts have widths of 6, 2 and
1\,$\mathrm{\mu}$m, with edge-to-edge gaps of 3 and 4.5$\,\mathrm{\mu}$m, respectively. Further device details are
described in \cite{Salis2009}. A current is applied flowing from the injector 2 to the 100$\, \mu$m distant contact 1.
A non-local voltage $V_{NL}$ is measured between contacts 3 and 5 outside the charge current path, see
Fig.\,\ref{fig:1}(a). $V_{NL}$ is detected by standard lock-in techniques using a small ac-modulation $I_{AC}$ on top
of a dc-injection current $I_{DC}$. The measurements are performed in a dilution refrigerator equipped with a
home-built 3-axis vector magnet, allowing us to determine the magnetization direction of the iron bars to better than
$\mathrm{1°}$ by rotating the magnetic field during continued spin-valve measurements.

Electron spin polarization pointing along the Fe easy-axis $\mathbf{\hat{x}}$ is injected into the semiconductor below
contact 2 \cite{Zhu2001,Lou2007}, diffuses away and can be detected at contact 3 (the electron spin diffusion length
exceeds the detector distance \cite{Salis2009}). DNP can easily be produced in presence of non-zero $I_{DC}$
\cite{Kawakami2001,Strand2003,Salis2009}, where the electron spins are imprinted onto the nuclear spins via flip-flops.
The nuclear spin polarization then acts back on the electron spins as an effective Overhauser field $B_{N}$
\cite{Overhauser1953} causing electron spin precession. In a perpendicular field $B_Z$, the electron spins precess,
diffuse and dephase, giving a characteristic Hanle peak around $B_Z=0$ \cite{Johnson1985, Jedema2002, Lou2007}. For
$\mathbf{B_N}$ antiparallel to $\mathbf{B}$, additional satellite peaks, see Fig.\,\ref{fig:1}(b), appear
\cite{Salis2009} when dephasing is suppressed by a cancelation of the external field by the internal Overhauser field:
$\mathbf B = -\mathbf B_N $. In the following, we will use this well established signature as a sensitive measure for
the nuclear field $B_N$ \cite{Meier,Farah1998,Epstein2002,Salis2009}. Nuclear fields achieved are $\sim 50\, $mT,
roughly one percent of the $5.3\, $T for fully polarized nuclei in GaAs \cite{Paget1977}. The average nuclear field
$B_N$ in our experiments is linear in $B_X$ \cite{Salis2009} along the Fe bars and for the following we fix $B_{X} =
-1.5\, $mT.

\begin{figure}[tpb]
\includegraphics[width=9cm]{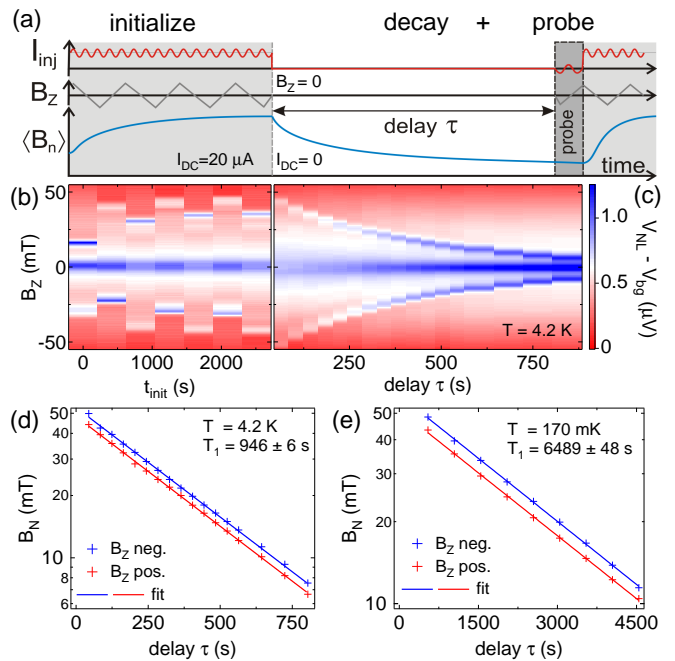}
\caption{\label{fig:2}(a) Pump-probe scheme used to measure the nuclear spin relaxation rate. (b) Initialization:
Alternating $B_Z$ Hanle sweeps (0.3\,mT/s) with $I_{DC}=20\,\mathrm{\mu A}$, see text. Sweeps start at $B_Z=+75\,$mT
and then run between $B_Z=\pm75\,$mT. (c) $B_Z$ probe-traces (0.9\,mT/s) after a delay $\tau$. Time-decay of the
satellites is clearly visible. A parabolic background was subtracted (same for all $\tau$). (d) and (e) Log-plot of
Overhauser field $B_{N}$ (crosses) -- extracted from satellite peak positions such as in (c) -- as a function of $\tau$
at 4.2\,K in (d) and 170\, mK in (e). Blue data is from satellites at $B_Z<0$, red from $B_Z>0$.
\emph{Single}-exponential fits (solid lines) give excellent agreement, and long $T_1$ times characteristic of NSR.}
\end{figure}

The pump-probe cycle used to find the NSR times is sketched in Fig.\,\ref{fig:2}(a). First, a nuclear polarization is
built-up by DNP while continuously sweeping $B_{Z}$ back and forth (`initialize'), see Fig.\,\ref{fig:2}(b), until a
steady state is reached, typically after an hour. The asymmetry, alternating positions and alternating widths of the
satellite peaks are a consequence of ramping and alternating sweep directions (DNP is most efficient at $B_Z\sim 0$
followed by slow decay at $B_Z\neq 0$ during ramping). After initialization, DNP is switched off ($I_{DC,AC}=0$) and
$B_{Z}$ is ramped to zero. The nuclear polarization is then allowed to decay for a time $\tau$ (`decay'), keeping $B_X
= -1.5\,$mT fixed. Subsequently, a fast Hanle scan to read out $B_N$ is performed (`probe') with only a small $I_{AC}$
and $I_{DC}=0$ to avoid further DNP during probing.

Repeating this cycle for various delays $\tau$ (including reinitializing each time), data sets reflecting the decay of
$B_N$ over time are obtained, as shown in Fig.\,\ref{fig:2}(c). By fitting Lorentzians to the satellite peaks, we
determine $B_N$ (peak position) as a function of $\tau$, as shown in Fig.\,\ref{fig:2}(d)/(e) (crosses), for both
positive (red) and negative (blue) $B_Z$ satellites. The small difference between the two satellite positions is a
result of slow ramping. From single-exponential fits, we get excellent agreement with the data, and $T_1$ times which
are the same within the error bars for the two satellites. Further, we observe sharpening of the satellites with
growing $\tau$, indicating increasing homogeneity of the nuclear spins with time. At temperatures above $1\,$K, the
$T_1$ times obtained here are in good agreement with previous $T_1$ measurements by NMR for all three isotopes
($^{69}$Ga,$^{71}$Ga,$^{75}$As) at comparable charge density \cite{Lu2006,Kaur2010}.

\begin{figure}[tpb]
\includegraphics[width=8.5cm]{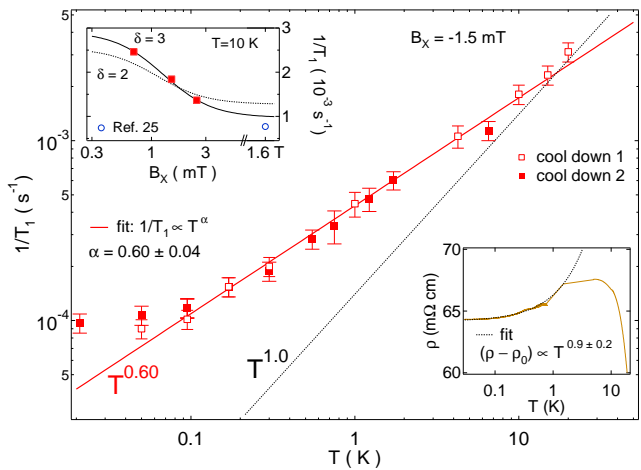}
\caption{\label{fig:3}The nuclear spin relaxation rate $1/T_1$ versus temperature measured for two cool downs (open and
solid squares) on the same sample, always for $B_X = -1.5\,$mT. Error bars are from repeated measurements. The solid
line is a power-law fit $1/T_1 \propto T^{\alpha}$ giving $\alpha=0.6\pm0.04$ for $\mathrm{0.1\,K}\leq
T\leq\mathrm{10\,K}$. As a comparison, an estimated Korringa law $1/T_{1} \propto T$ is added (dashed line) based on
NMR data \cite{Lu2006}, see text. Upper inset: $B_X$ dependence of the nuclear $T_{1}$-rate at $T=10\,\mathrm{K}$ with
theory (black curves, eq.\,(\ref{eq:3})), see text. An NMR data point at $B=1.6\,T$ and $T=10\,K$ from Ref.
\cite{Lu2006} is also added (rescaled using $1/T_1\propto n^{2/3}$ to match the carrier density here), demonstrating
very good agreement with the present spin-valve data. Lower inset: $T$-dependence of the resistivity from van der Pauw
measurements on the same GaAs wafer, indicating metallic behavior for $T<10\,$K. Dashed curve is a fit for $0.1\,K\leq
T\leq 1$K to $(\rho(T)-\rho_0)\propto T^\gamma$ giving $\gamma=0.9\pm0.2$.}
\end{figure}

The temperature dependence of the NSR rate is shown in Fig.\,\ref{fig:3} on a log-log plot for two cool downs (open and
closed squares) of the same sample. Measurements of a second sample (not shown) fabricated from another part of the
same wafer give very similar results. Both ac and dc currents were chosen to avoid self-heating over the measured
$T$-range. However, in the refrigerator used, sample temperatures saturate around $100\, $mK due to poor
thermalization, causing the relaxation rates to saturate at $100\, $mK. Nevertheless, at the lowest temperatures, very
long $T_1$ times exceeding 3 hours are found. Since the NSR rate in the log-log plot is linear over two orders of
magnitude in $T$, we fit a power law $1/T_{1} \propto T^{\alpha}$ for  $0.1\,$K$\,\leq T \leq 10\,$K and find
$\alpha=0.6 \pm 0.04$. The data at $T>10\,$K is excluded from the fit due to well known phonon contributions
\cite{McNeil1976}. For comparison, the Korringa law extrapolated from NMR data at $T>1\,K$ \cite{Lu2006} is indicated
in Fig.\,\ref{fig:3} (dashed line), rescaled from high density $n=2\times10^{18}\,\mathrm{cm^{-3}}$ where the Korringa
law holds to match the density in our samples using $1/T_1\propto n^{2/3}$ (Korringa scaling) and field corrected from
$1.6\,$T (NMR data) to $1.5$\,mT with a factor of 1.9 (see upper inset in Fig.\,\ref{fig:3}). The Korringa
$T$-dependence is clearly inconsistent with our data, which decreases more weakly with $T$ and gives relatively fast
NSR rates at low-$T$.
%We note that a similar (sub-linear)
%$T$-dependence above 1\,K seems to have been measured at $B=$1.6\,T with NMR in Ref. \cite{Lu2006} (though it was not
%discussed) -- suggesting that our finding is not a low-field ($B_X\,=$\,-1.5\,mT) peculiarity.

We now discuss the possible mechanisms of NSR. First, we exclude phonon contributions since these have been shown to be
relevant only well above $10\,$K and further would result in a quadratic temperature dependence
\cite{Lu2006,McNeil1976}. Also, NSR by paramagnetic impurities is known to be very weak in GaAs \cite{Lu2006}. Next, we
consider nuclear spin diffusion out of the $1\,\mathrm{\mu m}$ thick epilayer. This random-walk process is in principle
temperature independent in the regime applicable here and is inconsistent with the clear single-exponential decay of
$B_N(\tau)$ which we find for all temperatures, also making it unlikely that the observed low-$T$ saturation of
$T_1^{-1}$ is caused by nuclear spin diffusion. Therefore, we can exclude diffusion alone as a relevant relaxation
channel.

Next, we consider the hyperfine Fermi contact interaction as a possible NSR mechanism. In non-degenerate
semiconductors, where the Fermi energy is well below the conduction-band edge, the mobile charge carriers follow a
Boltzmann distribution, and the nuclear spin relaxation rate is $T_1^{-1}\propto \sqrt{T}$ \cite{Abragam}, not far from
the measured $T_1^{-1}\propto T^{0.6}$. However, since here $E_{F}\gg k_{B}T$ and since the measured resistivity
$\rho(T)$ in the relevant temperature range $T<10\,$K does not display a thermally activated behavior expected for a
non-degenerate semiconductor (see lower inset of Fig.\,\ref{fig:3}), this mechanism is most likely not applicable here.
In simple metals and degenerate semiconductors, the Korringa law is expected \cite{Korringa1950,Abragam}
\begin{equation}
\frac{1}{T_{1}} = \frac{256\pi^3}{9\hbar} \frac{\gamma _{n}^{2}}{\gamma _{e}^{2}}\ n^2 |\phi(0)|^{4} \chi^{2}\cdot
k_{B}T \ , \label{eq:4}
\end{equation}
with gyromagnetic ratio $\gamma_n$ of the nuclei and $\gamma_e$ of the electrons, electron spin susceptibility $\chi$
and $n\,|\phi(0)|^{2}$ is the electron density at the nuclear site. Indeed, this temperature dependence is observed in
much more highly-doped bulk GaAs ($n = 2\times 10^{18}\,$cm$^{-3}$) \cite{Lu2006} measured with NMR above $1\, $K, but
is not seen in the present samples.

To learn more about the mechanism of NSR present here, we investigate the $B_{X}$ dependence of $T_1^{-1}$, shown in
the upper inset of Fig.\,\ref{fig:3} at $10\, $K. Note that $B_Z=0$ during the decay step of the $T_1$ measurement. A
clear reduction of relaxation rates is seen for increasing $B_X$, as expected for applied fields comparable with $B_L$,
which is the local rms field acting on each individual nuclear spin, including nuclear dipole-dipole fields $B_d$ and
electronic Knight fields. The theoretically expected rate is \cite{Anderson1959, Hebel1959}
\begin{equation}
T_{1}^{-1}(B) = a \frac{B^2+\delta(5/3) B_{L}^2}{B^2+(5/3) B_{L}^2} \ , \label{eq:3}
\end{equation}
with large-field rate $a=T_{1}^{-1}(B\gg B_L)$. Note that the zero-field rate $T_1^{-1}(B=0)=\delta a$ and the
correlation parameter $\delta$ is ranging from 2 for uncorrelated to 3 for fully spatially correlated fields $B_L$.
Independent measurements give a very small $B$-field offset $<0.1\,\mathrm{mT}$, which we assume to be zero here. We
perform a fit and obtain $\delta = 3.0 \pm 0.3$, $B_L = 1 \pm 0.2\,$mT and $a=(9.6\pm1.5)\times
\,10^{-4}\,\mathrm{s^{-1}}$. The dashed curve shows a best-fit with $\delta = 2$, clearly inconsistent with the present
data. Taking the $B$-dependence from $\delta=3$ theory (solid curve, upper inset Fig.\,\ref{fig:3}), this brings the
spin-valve NSR rate at $B_X=-1.5$\,mT into very good agreement with NMR data measured at $B\sim1.6\,$ T and the same
$T=10$\,K \cite{Lu2006} (blue circle). Since $\delta=3$, $B_L$ is spatially highly correlated with a local field $B_L$
much larger than the estimated $B_d\sim 0.1\,$mT \cite{Paget1977} alone. This suggests an electronically induced
hyperfine mechanism causing NSR, due to electrons extended on a length scale much larger than the lattice constant
$a_0=5.7\,$\AA.

Since NSR appears to be electron mediated, we now discuss electronic transport measurements characterizing the
epilayer. The lower inset of Fig.\,\ref{fig:3} shows $\rho(T)$ from van der Pauw measurements done on separate samples
from the same wafer. Clearly, metallic behavior ($d\rho/dT\,>\,\mathrm{0}$) is seen for $T\,<\,\mathrm{10\,K}$, as
expected for the present doping of $\mathrm{5\, \times\, 10^{16}\,cm^{-3}}$, well above the well-known MIT in GaAs at
$\mathrm{n_c \sim 2 \times 10^{16}\,cm^{-3}}$ \cite{Rentzsch1986}. However, $\rho(T)$ is only weakly $T$-dependent
below 4\,K and follows $(\rho(T)-\rho_0)\propto T^{0.9\pm0.2}$ for $0.1\,\mathrm{K}\leq T \leq 1\,\mathrm{K}$,
deviating from the expected $\propto -T^{1/2}$ for the weak localization and Altshuler-Aronov corrections in 3D
\cite{AAreview}. We note that the simple Fermi liquid (FL) $\propto T^2$ is not expected here \cite{Maslov2012}. Above
10\,K, $\rho(T)$ shows simple thermal activation of donors \cite{Lu2006}. The carrier density at 4\,K is the same as at
base temperature (within measurement error), therefore excluding significant $T$-dependent carrier localization below
4\,K. Further, a perpendicular magnetic field has no significant effect for $\mathrm{B\,<\,5\,T}$ and gives a positive
magnetoresistance at larger fields. Therefore, the resistivity data shows clear metallic behavior, lacking any hints of
incipient localization.

In addition, control experiments have confirmed that the highly-doped surface layer does not significantly contribute
to lateral transport apart from facilitating the spin injection. The interaction parameter $r_{S} = E_{C}/E_{F}$ is
about 0.6, with Fermi energy $E_{F}\,=$\,7.4\,meV and average Coulomb energy $E_{C}\,=\,$4.1\,meV, indicating that the
samples are approaching the interacting regime $r_S\,\gtrsim$\,1. Further, disorder is quite strong: $k_F \ell
\sim$\,1.7, with a transport mean free path $\ell =$\,15\,nm for $T\,<$\,10\,K. Therefore, the epilayer behaves like a
degenerately doped semiconductor showing clear metallic behavior, in the interacting and strongly disordered regime.

Returning now to the NSR mechanism, the Korringa formula eq.\,(\ref{eq:4}) (where free electrons were assumed) would
need to be properly recalculated, including the combined effects of disorder and interactions not far from the MIT. In
lack of an appropriate theory in this regime, naively, a renormalized, temperature dependent electron spin
susceptibility $\chi(T)$ can be introduced in eq.\,(\ref{eq:4})
\cite{AAreview,Shastry1994,Fulde1968,Finkelshtein1984,Castellani1986,Belitz1991,Lee2000}. Here, $\chi\propto
T^{-\beta}$ with $\beta=\mathrm{0.2\pm0.02}$ would be required to result in $T_1^{-1}\propto T^{0.6}$ as measured,
assuming no other T dependencies in eq.\,(\ref{eq:4}). While $\beta=0$ corresponds to a regular FL, $\beta=$\,0.2 is in
good agreement with expectations ($0<\beta<1$) for the regime often associated with coexistence of localized moments
and itinerant electron states well within the metallic density range \cite{Miranda2005,CastroNeto1998}. Also, such a
low-temperature divergence of the spin susceptibility $\chi\propto T^{-\beta}$ has been observed in other
semiconductors for $n\gtrsim n_c$ above but not far from the MIT \cite{Paalanen1985,Sarachik1985,Manyala2008}. The
density dependence of $\beta$ would be interesting to investigate, indeed, but is beyond the scope of this study.

\begin{figure}[tpb]
\includegraphics[width=8.5cm]{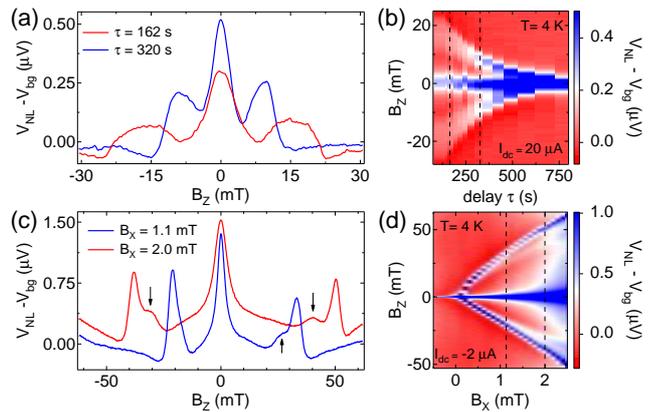}
\caption{\label{fig:4}(a) and (b) are Hanle probe sweeps with $B_Z = 0$ during initialization, showing broadened Hanle
peaks but resulting in very similar NSR rates (not shown). (c) and (d): Double satellite peaks in slow Hanle
measurements ($0.25\,$mT/s) become visible (arrows in (c)) for $B_{X}>0.8\,$mT.}
\end{figure}

Finally, we investigate nuclear spin inhomogeneities apparent in the Hanle measurements. When fixing
$\mathrm{B_Z\,=\,0}$ during initialization, significantly broadened satellite peaks result, see
Fig.\,\ref{fig:4}(a)/(b), though the extracted NSR rates remain unchanged within experimental error (not shown). As
seen by comparing Fig.\,\ref{fig:2}(c) with Fig.\,\ref{fig:4}(b), sweeping $B_Z$ (during initialization) has the effect
to narrow the Hanle peaks, apparently homogenizing the nuclear spins. Further, we find additional satellite peaks, see
Fig.\,\ref{fig:4}(c)/(d), suggesting two distinct species of electrons and/or nuclear polarization regions. We note
that the extra satellites are visible whenever they are sufficiently sharp and well-enough separated, independent of
the current direction and the sign of $B_X$. Further studies are needed to elucidate these additional satellite peaks
as well as inhomogeneity effects.

In summary, using a new, versatile method to measure NSR in spin-valve devices, we report the breakdown of the
Korringa-law in GaAs doped a factor of $\sim2.5$ above the MIT displaying clearly metallic conductivity. Over a factor
of $100$ in $T$, the NSR rate follows a rather weak power-law $1/T_1\propto T^{0.6}$, resulting in relatively strong
coupling and NSR rates enhanced beyond the Korringa law at low-$T$, potentially useful for nuclear cooling. This
power-law is consistent with a weakly diverging electron spin susceptibility $\chi\propto T^{-0.2}$ in the
simultaneously interacting and disordered metallic-regime not far from the MIT currently lacking appropriate theory.

We are very thankful for discussions with B. Braunecker, D. Loss, D. Maslov, and S. Valenzuela. This work was supported
by the Swiss Nanoscience Institute (SNI), Swiss NSF, NCCR NANO, NCCR QSIT and an ERC starting grant (DMZ).

\end{document}